\documentstyle[12pt]{article}
\begin{document}
\centerline{\Large \bf Wormholes in String Theory}
\vskip .9in
\centerline{Dan N. Vollick}
\centerline{Department of Physics and Astronomy}
\centerline{University of Victoria}
\centerline{Victoria, British Columbia}
\centerline{P.O. BOX 3055 MS7700}
\centerline{Canada}
\centerline{V8W 3P6}
\vskip .9in
\section*{\centerline{\bf Abstract}}
A wormhole is constructed by cutting and joining two spacetimes satisfying
the low energy string equations with a dilaton field. In spacetimes described
by the ``string metric"
the dilaton energy-momentum tensor need not satisfy the weak or dominant
energy conditions. In the cases considered here the dilaton field violates
these energy conditions and is the source of the exotic matter required
to maintain the wormhole. There is also a surface stress-energy, that must be
produced by additional matter, where the spacetimes are joined. It is
shown that wormholes can be constructed for which this additional matter
satisfies the weak and dominant energy conditions, so that it
could be a form of
``normal" matter. Charged dilaton wormholes with a coupling
between the dilaton and the electromagnetic field that is more general than
in string theory are also briefly discussed.
\section*{Introduction}
To keep a wormhole open it is necessary to thread its throat with 
matter that violates the averaged null energy condition 
(also called the averaged weak energy condition) [1-6]. 
That is, there must exist null geodesics with tangent vectors $k^{\mu}=dx^{\mu}
/ds$ that satisfy $\int_0^{\infty}T_{\mu\nu}k^{\mu}k^{\nu}ds<0$. 
Most discussions of such
exotic matter involve quantum field theory effects, such as the Casimir effect
\cite{It1} (see \cite{Vo1,Vo2} for a classical discussion). In a recent
paper Ford and Roman \cite{Fo1} have shown that the quantum inequalities
satisfied by the negative energy densities in scalar and vector quantum
field theories tightly constrain the geometry of wormholes. Their analysis 
shows that the wormhole can either be on the order of $\simeq$ Planck size
or that the negative energy density is concentrated in a ``thin-shell"
at the throat of the wormhole. In another recent paper Taylor, Hiscock,
and Anderson \cite{Ta1}
have examined the energy-momentum tensor of a ``test" quantized
scalar field in a fixed background wormhole spacetime. They found that
for five different wormhole geometries the energy-momentum tensor of a
minimally or conformally coupled scalar field does not even come close to
having the properties required to support the wormhole. Thus at present,
it appears that the prospect of maintaining a wormhole through quantum
field theory effects is not very promising.
    
In this paper I
examine the possibility of using the dilaton field, that appears in string
theory, as the source of the exotic matter. Solutions to low energy string
are found that have a throat connecting an asymptotically flat spacetime
to one that has a curvature singularity at infinity. 
This singularity can be removed, however, by adding matter to the spacetime.
In the wormholes examined here this additional matter takes the form of 
a thin spherical shell.
To construct a wormhole I take two
static, spherically symmetric spacetimes satisfying the low energy string
equations, cut out the singular regions 
and join them together. The dilaton field turns out to violate the
weak and dominant energy conditions, and thus serves as the exotic matter
required to maintain the wormhole. There is also a surface distribution
of stress energy induced where the spacetimes are joined. I
show that this matter satisfies the weak and dominant energy conditions,
so that it could be a form of ``normal" matter. Of course, the additional
matter needed to eliminate the curvature singularity need not be in the form
of a thin shell, it could have been spread out in space.

I also briefly examine charged dilaton wormholes in which the coupling of
the dilaton to the electromagnetic field is generalized from the string case.
\section*{Field Equations and Spherically Symmetric Solutions}
The action for gravity coupled to a dilaton and Maxwell field will be 
taken to be
\begin{equation}
S=\int d^4x\sqrt{g}\left\{ e^{-2\phi}[R-4\nabla_{\mu}\phi\nabla^{\mu}\phi]
+e^{-2a\phi}
F_{\mu\nu}F^{\mu\nu}\right\} ,
\label{action}
\end{equation}
where $R$ is the Ricci scalar, $\phi$ is the dilaton field, and $F_{\mu\nu}$
is the electromagnetic field tensor. For $a$=0 the action reduces to the
usual Einstein-Maxwell-scalar theory and for $a=1$ the action becomes the
low energy string action with the antisymmetic field set to zero.
  
The field equations that follow from (\ref{action}) are
\begin{equation}
\nabla_{\mu}\left ( e^{-2a\phi}F^{\mu\nu}\right )=0,
\label{F}
\end{equation}
\begin{equation}
\Box^2\phi-\nabla_{\mu}\phi\nabla^{\mu}\phi-\frac{1}{4}R-\frac{a}{4}e^{2(1-a)
\phi}F_{\mu\nu}F^{\mu\nu}=0
\label{Phi}
\end{equation}
and,
\begin{equation}
R_{\mu\nu}=2\nabla_{\mu}\nabla_{\nu}\phi-2e^{2(1-a)\phi}\left [F_{\mu\alpha}
F_{\nu}^{\;\;\alpha}-\frac{1}{4}(1-a)g_{\mu\nu}F^{\alpha\beta}F_{\alpha\beta}
\right ],
\label{R}
\end{equation}
where I have used the $\phi$ field equation to simplify equation (\ref{R}).
The static spherically symmetric solutions to the above field equations can be 
written as \cite{Ga1,Ho1}
\begin{equation}
ds^2=-\lambda^2 dt^2+\gamma^2 dr^2+R^{*2}d\Omega^2 ,
\label{metric}
\end{equation}
\begin{equation}
e^{2(\phi-\phi_0)}=\left(1+\frac{B}{r}\right )^{\frac{2a}{1+a^2}} ,
\end{equation}
\begin{equation}
F_{tr}=\frac{Q}{r^2} ,
\end{equation}
where,
\begin{equation}
\lambda^2=\left( 1+\frac{A}{r}\right)\left(1+\frac{B}{r}\right)^{
\frac{1+2a-a^2}{1+a^2}} ,
\label{lambda}
\end{equation}
\begin{equation}
\gamma^2=\left( 1+\frac{A}{r}\right)^{-1}\left(1+\frac{B}{r}\right)^{
\frac{a^2+2a-1}{1+a^2}} ,
\end{equation}
\begin{equation}
R^*=r\left(1+\frac{B}{r}\right)^{\frac{a^2+a}{1+a^2}} ,
\label{radius}
\end{equation}
$A$ and $B$ are two independent parameters, and $\phi_0$ is the asymptotic
value of $\phi$. The mass and charge are given by
\begin{equation}
M=\frac{B}{2}\left(\frac{a^2-2a-1}{1+a^2}\right)-\frac{1}{2}A
\label{M}
\end{equation}
\begin{equation}
Q=\left[\frac{AB}{1+a^2}\right]^{1/2}e^{(1-a)\phi_0} .
\label{Q}
\end{equation}
In the black hole case $A$ is taken to be negative, so that an event horizon
exists for sufficiently small $Q$
(note that this implies that $B<0$ for $Q$ to be real). Here, I am
interested in wormholes without horizons, so I will take $A,B>0$. This is
sufficient but not necessary for the absence horizons, since for large
enough $Q$ the horizon disappears even if $A,B<0$. 
However, the spatial sections will
have a throat iff $B>0$ and $a>1$. 
To see this consider equation (\ref{radius}) with
$B>0$ and $a>1$. For large $r$ the surface area of a sphere decreases with
decreasing $r$ until
\begin{equation}
r_{th}=\left(\frac{a-1}{1+a^2}\right)B ,
\label{throat}
\end{equation}
after which it increases with decreasing $r$. Thus the spacetime contains a 
wormhole with a throat at $r_{th}$. Since it is necessary that
$a>1$ for a wormhole throat to exist it may seem that wormholes cannot
be constructed in string theory by this method.
However, this is not the case if $Q=0$, since $a$ disappears from the 
action and is in fact a free parameter in the solution.
Unfortunately, the spacetime also contains
a curvature singularity at $r=0$ in addition to a throat. 
There are other forms of matter, however,
that satisfy the weak and dominant energy conditions, at our disposal. The 
question then becomes: can we distribute such matter in the spacetime
eliminating the singularity but maintaining the wormhole structure? In the next
section I will construct nonsingular wormholes by cutting and pasting two
spacetimes satisfying equations (\ref{metric}) to (\ref{Q}) with $Q=0$.
One manifold $M^+$ will consist of the region $r_+
\leq r < \infty$ with $r_+\geq r_{th}$, and the other manifold $M^-$
will consist of the region $r_-\leq r < \infty$ with $r_-\leq r_{th}$. Note
that $R^*(r_-)=R^*(r_+)$ must be satisfied.
Additional matter, that satisfies the weak and dominant energy
conditions, exists on the surface where the two manifolds
are joined.
\section*{String Wormholes}
Consider taking $A=0$ in equations (\ref{metric})-(\ref{Q}). 
This implies that $Q=0$ and 
$F_{\mu\nu}=0$. The action (\ref{action}) becomes independent of $a$ and
the solution (\ref{metric}) to (\ref{Q}) is still the static, spherically
symmetric solution with $a$ now being a free parameter. It is well-known that
the static, spherically symmetric solution to the usual Einstein-scalar
field equations contains an additional free parameter related to the scalar
charge. Thus, in string theory a wormhole throat can be created by taking
$A=0$ and $a>1$. The mass of the wormhole is given by
\begin{equation}
M=\frac{1}{2}\left(\frac{a^2-2a-1}{a^2+1}\right) ,
\end{equation}
which is negative for $1<a<1+\sqrt{2}$ and positive for $a>1+\sqrt{2}$.
The Ricci scalar is given by
\begin{equation}
R=2\Box^2\phi=\frac{4a^2B^2}{(1+a^2)^2r^4}\left(1+\frac{B}{r}\right)^
{-\left(\frac{3a^2+2a+1}{1+a^2}\right)} ,
\end{equation}
so that there is a curvature singularity at $r=0$ for all $a\neq 0$. 
The energy-momentum tensor 
of the dilaton field is
\begin{equation}
T_{\mu\nu}=-\frac{1}{4\pi G}\left(\nabla_{\mu}\nabla_{\nu}\phi-\frac{1}{2}
g_{\mu\nu}\Box^2\phi\right) 
\end{equation}
and the corresponding energy density $\rho=-T^t_{\;\; t}$ is given by
\begin{equation}
\rho=-\frac{a(a^2-1)}{8\pi G(1+a^2)^2r^4}\left(1+\frac{B}{r}\right)^{-
\left(\frac{3a^2+2a+1}{1+a^2}\right)} .
\end{equation}
This obviously violates the weak and dominant energy conditions for $a>1$.
Note that as $r\rightarrow 0$ the
energy density diverges for all $a\neq 0,\pm 1$.
  
As discussed in the introduction I will join two manifolds, $M^+$ consisting
of the region $r_+\leq r<\infty$ and $M^-$ consisting of the region
$r_-\leq r<\infty$, together. 
The coordinate radii $r_+$ and $r_-$ will be taken
to satisfy $r_-\leq r_{th}\leq r_+$ and $R^*(r_+)=R^*(r_-)$. Before joining the
two manifolds together it will be convenient to change coordinates. Let $l$ be
a new radial coordinate with $l=0$ at $r=r_{th}$ and
\begin{equation}
dl^2=\left(1+\frac{B}{r}\right)^{\frac{a^2+2a-1}{1+a^2}}dr^2 .
\end{equation}
The joined manifold will have the metric
\begin{equation}
ds^2=-\eta^2_{\pm}\left(1+\frac{B}{r}\right)^{\frac{1+2a-a^2}{1+a^2}}dt^2+dl^2+
r^2\left(1+\frac{B}{r}\right)^{\frac{2(a^2+a)}{1+a^2}}d\Omega^2
\end{equation}
where $r=r(l)$, $-\infty <l<\infty$, and $l\rightarrow\pm\infty$ as 
$r\rightarrow\infty$ on $M^{\pm}$. Note that two constants $\eta_+$ and
$\eta_-$ have been introduced by redefining the time coordinate. These
constants will be used to make $g_{tt}$ continuous across the join. The 
coordinates $r$ and $l$ are related via
\begin{equation}
\frac{dr}{dl}=\pm\left(1+\frac{B}{r}\right)^{\frac{1-2a-a^2}{2(1+a^2)}}
\end{equation}
with the upper sign on $M^+$ and the lower sign on $M^-$.
  
The surface energy-momentum tensor 
\begin{equation}
S^{\mu}_{\;\;\nu}=\lim_{\epsilon\rightarrow 0}\int_{-\epsilon}^{\epsilon}
T^{\mu}_{\;\;\nu}dl
\end{equation}
is given by [13-15]
\begin{equation}
8\pi GS^{\mu}_{\;\;\nu}=\gamma^{\mu}_{\;\;\nu}-\delta^{\mu}_{\;\;\nu}
\gamma\;\;\;\;\;\;\;\;\;\; (\mu,\nu=0,2,3)
\end{equation}
where
\begin{equation}
\gamma^{\mu}_{\;\;\nu}=K^{+\mu}_{\;\;\;\;\nu}-K^{-\mu}_{\;\;\;\;\nu} ,
\end{equation}
and $K^{\pm\mu}_{\;\;\nu}$ is the extrinsic curvature on $M^{\pm}$ at $r_{\pm}$.
Using $K_{\mu\nu}=-\frac{1}{2}g_{\mu\nu ,l}$ gives
\begin{equation}
S^t_{\;\; t}=\frac{1}{4\pi G}\left[\frac{(r_+-r_{th})}{r_+^2}\left(1+\frac{B}
{r_+}\right)^{-\left(\frac{3a^2+2a+1}{2(1+a^2)}\right)}
+\frac{(r_--r_{th})}{r_-^2}\left(
1+\frac{B}{r_-}\right)^{-\left(\frac{3a^2+2a+1}{2(1+a^2)}\right)}\right]
\end{equation}
and
\begin{equation}
\begin{array}{ll}
S^{\theta}_{\;\;\theta}=S^{\phi}_{\;\;\phi}=\frac{1}{8\pi G}\left[\frac{1}{r_+}
\left(1+\frac{(a^2-4a+1)B}{2(1+a^2)r_+}\right)\left(1+\frac{B}{r_+}\right)^
{-\left(\frac{3a^2+2a+1}{2(1+a^2)}\right)}\right.\\
\left. +\frac{1}{r_-}\left(1+\frac{(a^2-4a+1)B}{2
(1+a^2)r_-}\right)\left(1+\frac{B}{r_-}\right)^{-\left(
\frac{3a^2+2a+1}{2(1+a^2)}\right)}
\right]
\end{array}
\end{equation}
The surface energy density is $\sigma=-S^t_{\;\;t}$ and the surface pressure
is given by $P=S^{\theta}_{\;\;\theta}$. 
  
Consider the limit $r_+>>r_{th}$ and $r_-<<r_{th}$. In this limit
\begin{equation}
\sigma=\frac{r_{th}}{4\pi Gr_-^2}\left(\frac{r_-}{B}\right)^
{\frac{3a^2+2a+1}{2(1+a^2)}}>0
\end{equation}
and
\begin{equation}
P=\frac{B}{16\pi Gr_-^2}\left(\frac{r_-}{B}\right)^{\frac{3a^2+2a+1}{2(1+a^2)}}
\left(\frac{a^2-4a+1}{1+a^2}\right).
\end{equation}
The weak energy condition will hold if $\sigma\geq 0$ and $\sigma+P\geq 0$.
The first condition is automatically satisfied and the second condition will
be satisfied if $a\geq\sqrt{3}$. The dominant energy condition holds if
$\sigma\geq |P|$. This will be satisfied if $\sqrt{3}\leq
a\leq 4+\sqrt{11}$. And finally, the strong energy condition will hold if
$\sigma +P\geq 0$ and $\sigma +2P\geq 0$, which implies that $a\geq 1+\sqrt{2}
$. The weak and dominant energy conditions follow from the positivity of $\rho$
and from the non-spacelike nature of the energy flow vector, respectively.
The strong energy condition arises only as a mathematical condition
in some of the singularity theorems. Thus
$a$ will be taken to satisfy $\sqrt{3}\leq a\leq 4+\sqrt{11}$.
      
To complete the description of the wormhole the source of the scalar field
must be examined. In the presence of a source $S$ the scalar field equation 
becomes
\begin{equation}
\Box^2\phi-\nabla_{\mu}\phi\nabla^{\mu}\phi-\frac{1}{4}R=4\pi S\delta(l-R_l)
\label{source}
\end{equation}
where $R_l$ is the radius of the sphere where the manifolds are joined
in the $l$ coordinate system. 
Requiring that the scalar field be continuous across the join gives
\begin{equation}
e^{2\phi_0^+}\left(1+\frac{B}{r_+}\right)^{\frac{2a}{1+a^2}}
=e^{2\phi_0^-}\left(1+\frac{B}{r_-}\right)^{\frac{2a}{1+a^2}}
\end{equation}
where $\phi_0^{\pm}$ is the asymptotic value of $\phi$ on $M^{\pm}$.
Integrating (\ref{source}) from $R_l-\epsilon$ to $R_l+\epsilon$ gives
\begin{equation}
\left[\sqrt{g}\frac{\partial\phi}{\partial l}\right]^{R_l+\epsilon}_{R_l
-\epsilon}-\int^{R_l+\epsilon}_{R_l-\epsilon}\left(\frac{\partial\phi}
{\partial l}\right)^2\sqrt{g}dl-\frac{1}{4}\int_{R_l-\epsilon}^{R_l+
\epsilon}R\sqrt{g}dl=4\pi\sqrt{g} S.
\label{jump}
\end{equation}
Since $\phi$ is continuous across $R_l$ the second term vanishes as $\epsilon
\rightarrow 0$. The Ricci scalar consists of terms containing the metric
and its second derivatives and terms containing the metric and its first
derivatives. Since the metric is continuous across $R_l$ only terms containing
the second derivatives of the metric will contribute in (\ref{jump}). $S$ is
well defined since
this term is independent of the metric's first derivatives and is
linear in its second derivatives.
     
Charged dilaton wormholes with $a>1$ can also be produced by similar methods.
In this case the charge associated with the two asymptotically flat regions
will be different and should be chosen so as to make the electromagnetic
field continuous across the jump. This implies that $A$ and $B$ will generally
have different values on $M^+$ and $M^-$.
\section*{Conclusion}
Solutions to the low energy string theory were found that have a throat
connecting an asymptotically flat universe to one that has a curvature
singularity at infinity. To eliminate the singularity I took
two of these manifolds, cut out the singular regions and joined them
together. The dilaton field violates the weak and dominant energy conditions
and is thus the source of the exotic matter required to hold the wormhole
open. The stress-energy on the surface where the manifolds were joined
satisfied the weak and dominant energy conditions and is therefore ``normal"
matter. I also briefly discussed creating charged 
dilaton wormholes in which the 
coupling of the dilaton to the electromagnetic field is more general
than in string theory.


\begin{thebibliography}{1}
\bibitem{Mo1}
M.S. Morris and K.S. Thorne, Am. J. Phys. {\bf 56}, 395 (1988)
\bibitem{Mo2}
M.S. Morris, K.S. Thorne, and U. Yurtserver, Phys. Rev. Lett. {\bf 61}, 1446 (1988)
\bibitem{Fr1}
V.P. Frolov and I.D. Novikov, Phys. Rev. D{\bf 42}, 1057 (1990)
\bibitem{Fr2} 
J.L. Friedman, K. Schleich, and D.M. Witt, Phys. Rev. Lett. {\bf 71}
1486 (1993) 
\bibitem{Vi1} 
M. Visser, \emph{Lorentzian Wormholes: From Einstein to Hawking} 
(AIP Press, 1996)
\bibitem{Hoc1}
D. Hochberg and M. Visser, Phys. Rev. D{\bf 56}, 4745 (1997)
\bibitem{It1}
C. Itzykson and J. Zuber, \emph{Quantum Field Theory} 
(McGraw-Hill, 1980), see sec. 3-2-4
\bibitem{Fo1}
L.H. Ford and T.A. Roman, Phys. Rev. D{\bf 53}, 5496 (1996)
\bibitem{Ta1}
B.E. Taylor, W.A. Hiscock, and P.R. Anderson, Phys. Rev. D{\bf 55}, 6116 (1997)
\bibitem{Vo1}
D.N. Vollick, Phys. Rev. D{\bf 56}, 4720 (1997)
\bibitem{Vo2}
D.N. Vollick, Phys. Rev. D{\bf 56}, 4724 (1997)
\bibitem{Ga1}
D. Garfinkle, G.T. Horowitz, and A. Strominger, Phys. Rev. D{\bf 43}, 3140 (1991)
\bibitem{Ho1}
J.H. Horne and G.T. Horowitz, Phys. Rev. D{\bf 46}, 1340 (1992)
\bibitem{Is1}
W. Israel, Nouvo Cimento B{\bf 44}, 1 (1966)
\bibitem{Is2}
W. Israel, Nouvo Cimento B{\bf 48}, 463 (1967)
\bibitem{Mi1}
C.W. Misner, K.S. Thorne, and J.A. Wheeler, \emph{Gravitation} (W.H. Freeman
an Company, 1973)
\end{thebibliography}
\end{document}